\documentclass[twocolumn,prl,superscriptaddress]{revtex4}

\usepackage{epsfig}
\usepackage{bm}

\newcommand{\qF}{\mathcal{F}}
\newcommand{\qT}{\mathcal{T}}
\newcommand{\qe}{\varepsilon}
\newcommand{\ra}{\rightarrow}

\newcommand{\qFn}{F_{\rm n}}
\newcommand{\qvel}{\mathbf{v}}
\newcommand{\qv}{v}
\newcommand{\qvv}{v^*}
\newcommand{\qomega}{\bm{\omega}}
\newcommand{\qo}{\omega}
\newcommand{\qoo}{\omega^*}

\begin{document}

\title{Frictional coupling between sliding and spinning motion}

\date{September 27, 2002}

\author{Z\'en\'o Farkas}
\author{Guido Bartels}
\affiliation{Inst.\ of Physics, Gerhard-Mercator University Duisburg,
D-47048 Duisburg, Germany}
\author{Tam\'as Unger}
\affiliation{Dept.\ of Theoretical Physics, Budapest University of Technology
and Economics, H-1111 Budapest, Hungary}
\author{Dietrich E. \surname{Wolf}}
\affiliation{Inst.\ of Physics, Gerhard-Mercator University Duisburg,
D-47048 Duisburg, Germany}

\begin{abstract}
We show that the friction force and torque, acting at a dry contact of
two objects moving and rotating relative to each other, are inherently
coupled. As a simple test system, a sliding and spinning disk on a
horizontal flat surface is considered. We calculate, and also measure,
how the disk is slowing down, and find that it always stops its
sliding and spinning motion at the same moment.
We discuss the impact of this coupling between friction force and
torque on the physics of granular materials.
\end{abstract}


\maketitle

Sliding friction and incomplete normal restitution are normally 
the main dissipation mechanisms at the contact between two solid
grains. They are largely responsible for the fact that the flow
properties of granular media differ from those for liquids and 
solids. Their microscopic origins are currently under intense
investigation (see e.g. \cite{Persson00}). On large scales compared to
the grain diameter they sometimes transform into unexpected
phenomenological friction laws, which have recently been discussed
\cite{Wolf98,Dippel99,RadjaiRoux95,Rajchenbach02,Ertas_et_al.01}.
As the influence of rolling and torsion friction is commonly regarded as 
negligible, they have been much less investigated so far. However, it
turns out that in certain situations they may become crucial, for
instance for the stabilization of pores in cohesive powders
\cite{Kadau_et_al.02,Kadau_et_al.03}. Another striking phenomenon will
be discussed below. 

In fact, rolling and torsion friction are indispensable for a
unified view of the dissipation mechanisms at the contact of two
viscoelastic spheres, because on the one hand incomplete restitution
and rolling friction, and on the other hand sliding and
torsion friction are coupled, which will be the main point of
this letter. These four dissipation mechanisms correspond to the 
six degrees of freedom of the relative motion at the contact between
two solid spheres. 
The relative motion of two solid spheres has three translational
degrees of freedom, characterized by a velocity vector with one
normal component $v_{\rm n}$ (deformation mode) and two tangential
components $v_{\rm t}$ (sliding mode), and three rotational ones,
characterized by an angular velocity vector, again with
two tangential components $\omega_{\rm t}$ (rolling mode) and one
normal component $\omega_{\rm n}$ (spinning mode).  
While the viscoelastic dissipation mechanism of normal restitution and
rolling friction couples $v_{\rm n}$ and $\omega_{\rm t}$
\cite{BrilliantovPoeschel98,Poeschel_et_al.99}, the dissipation due to
sliding and torsion friction couples $v_{\rm t}$ and $\omega_{\rm n}$. 

\begin{figure}
\centerline{\epsfig{figure=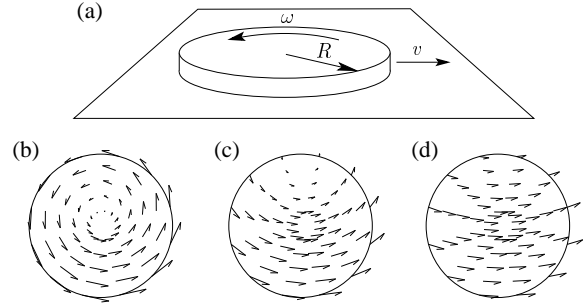,width=7.7cm}}
\caption{(a) A sliding and spinning disk on a flat horizontal
  surface.  (b)--(d) The relative velocity field on the surface of
  the disk at $\qe = 0.2$, $\qe = 1$, and $\qe = 5$, respectively
  ($\qe = \qv / R \qo$).}
\label{fig:DiskOnPlane}
\end{figure}

In this Letter we focus on the coupling between $v_{\rm t}$ and
$\omega_{\rm n}$. For viscoelastic spheres these are the dominant
dissipation channels in the quasistatic limit, where $(1- {\rm restitution
\; coefficient})$ \cite{KuwabaraKono87,Ramirez_et_al.99},
as well as the coefficient of rolling friction vanish
\cite{BrilliantovPoeschel98}. The reason, why torsion 
friction, i.e., the torque leading to a decrease of $\omega_{\rm n}$,  
is often neglected, is that it involves the radius of the contact
area between the two spheres and hence is small.
Therefore, in order to make our point more clear,
instead of the contact between two spheres
we consider a flat disk on a horizontal flat surface 
with nonzero initial translational and angular velocity.
The disk is lying on one of its sides, and we assume that
this side is in full contact with the table 
during the motion [see Fig.~\ref{fig:DiskOnPlane}(a)].
The friction force and torque acting on the disk will
slow down the sliding and spinning motion until the disk stops moving.
We address two questions: 
(i) how are the friction force and torque related to each other,
(ii) what does this imply for the coupling of sliding and spinning motion?
 
First we calculate the friction force and torque acting on the disk
as a function of its instantaneous velocity and angular velocity.
We apply the Coulomb friction law, which says that the magnitude 
of the friction force is proportional to the normal force, while
its direction is the opposite of the direction of the 
surfaces' relative velocity.
Assuming that the mass distribution of the disk is
homogeneous, the friction force is
$$
\mathbf{F} = - \frac{\mu \qFn }{R^2 \pi} \int\limits_{\mathbf{r}\in A}
\frac{\qvel + \qomega \times \mathbf{r}}
{|\qvel + \qomega \times \mathbf{r}|} d^2 r,
$$
where $R$ is the radius, $\qvel$ is the
velocity, and $\qomega$ is the angular velocity of the disk,
$\mu$ is the friction coefficient,
and the integration extends over the area of the disk with
$\mathbf{r}$ vectors starting at the center.
$\qFn$ is the normal component of the force pressing the objects
together at the contact, in our case
$\qFn = m g$, where $m$ is the mass of the disk and $g$ is the gravitational
acceleration.
We found it useful to introduce the dimensionless quantity
$\qe = \qv / R \qo$ with $\qv = |\qvel|$ and
$\qo = |\qomega|$, because
the friction force depends on $\qv$ and $\qo$
only through this combination:
\begin{equation}
\mathbf{F} = - \frac{\mu \qFn}{\pi} \int\limits_{\tilde\mathbf{r}\in A_1}
\frac{\qe \mathbf{e}_v + \mathbf{e}_\omega \times \tilde \mathbf{r}}
{|\qe \mathbf{e}_v + \mathbf{e}_\omega \times \tilde \mathbf{r}|}
 d^2 \tilde r,
\label{eq:force}
\end{equation}
where $\mathbf{e}_v = \qvel / \qv$,
$\mathbf{e}_\omega = \qomega / \qo$,
$\tilde \mathbf{r} = \mathbf{r} / R$, and $A_1$ is the area of the
unit disk. 
Figure \ref{fig:DiskOnPlane}(b)--(d) show local relative velocities
on the surface of the disk for various values of $\qe$.
Note that the local friction force does not depend on the absolute value
of the relative velocity, only on its direction.
After evaluating the integral in Eq.\ (\ref{eq:force}) one gets
$\mathbf{F} = - \mu \qFn \qF(\qe) \mathbf{e}_v$,
where 
$$
\qF(\qe) =
\left\{ \begin{array}{ll}
\displaystyle\frac{4}{3} 
\frac{
(\qe^2 + 1) E(\qe)
+ (\qe^2 - 1) K(\qe)}
{\qe \pi},
& \qe \le 1 \\
 & \\
\displaystyle\frac{4}{3} 
\frac{
(\qe^2 + 1) E(\frac{1}{\qe})
- (\qe^2 - 1) K(\frac{1}{\qe})}
{\pi},
& \qe \ge 1.
\end{array} \right.
$$
Here $K(\qe)$ and $E(\qe)$ are the complete
elliptic integral functions of the first and the second kind, respectively
\cite{abramowitz}.
This calculation and the others below were performed using
the mathematics software Maple \cite{maple}.
The two parts of $\qF(\qe)$ are smoothly connected at $\qe = 1$, since
$\lim_{\qe\ra 1} \qF(\qe) = 8 / 3 \pi$ and
$\lim_{\qe\ra 1} \qF'(\qe) = 4 / 3 \pi$
from both the left and the right hand side.
Here prime denotes differentiation w.r.t.\ $\qe$.
The limiting values are
$\qF(0) = 0$ and $\lim_{\qe\ra\infty} \qF(\qe) = 1$. 

\begin{figure}
\centerline{\epsfig{figure=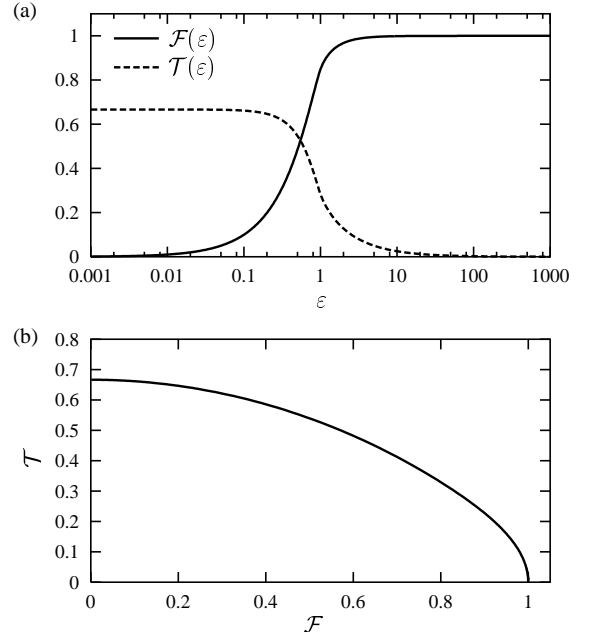,width=7.7cm}}
\caption{(a) The dimensionless friction force and torque,
$\qF$ and $\qT$, as functions of the dimensionless velocity parameter
$\qe$.
(b) The friction force and torque are coupled: the curve shows
the possible $(\qT,\qF)$ pairs.}
\label{fig:F_T_e}
\end{figure}

The friction torque is
$$
\mathbf{T} = - \frac{\mu \qFn }{R^2 \pi} \int\limits_{\mathbf{r}\in A}
\mathbf{r} \times \frac{\qvel + \qomega \times \mathbf{r}}
{|\qvel + \qomega \times \mathbf{r}|} d^2 r,
$$
and after calculating the integral we get
$\mathbf{T} = - \mu \qFn R \qT(\qe) \mathbf{e}_\omega$,
where 
$$
\qT(\qe) =
\left\{ \begin{array}{ll}
\displaystyle\frac{4}{9} 
\frac{
(4 - 2 \qe^2) E(\qe)
+ (\qe^2 - 1) K(\qe)}
{\pi},
& \qe \le 1 \\
 & \\
\displaystyle\frac{4}{9} 
\frac{
(4 - 2 \qe^2) E(\frac{1}{\qe})
+ (2 \qe^2 - 5  + \frac{3}{\qe^2}) K(\frac{1}{\qe})}
{\qe \pi},
& \qe \ge 1.
\end{array} \right.
$$
The two parts of this function are also smoothly connected, as
$\lim_{\qe\ra 1} \qT(\qe) = 8 / 9 \pi$ and
$\lim_{\qe\ra 1} \qT'(\qe) = -4/ 3 \pi$
from both the left and the right hand side.
The limiting values are
$\qT(0) = 2 / 3$ and $\lim_{\qe\ra\infty} \qT(\qe) = 0$.
Figure \ref{fig:F_T_e} shows $\qF(\qe)$ and $\qT(\qe)$,
and also the $\qT(\qF)$ function. This latter exists
and is invertable because both $\qF(\qe)$ and $\qT(\qe)$ are 
strictly monotonic functions.

\begin{figure}
\centerline{\epsfig{figure=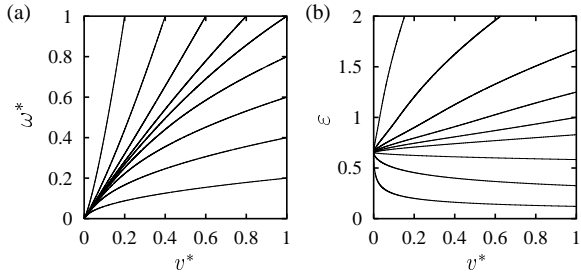,width=7.7cm}}
\caption{Numerical results: (a) $\qoo$--$\qvv$ trajectories with
different initial conditions.  (b) $\qe$--$\qvv$ trajectories with the
same initial conditions as in (a).}
\label{fig:trajectories}
\end{figure}

Now let us calculate how a sliding and spinning disk is slowing down.
Assuming that only gravity and friction forces are acting, the
scalar equations of motion are
\begin{eqnarray}
m \frac{d \qv}{d t} & = & - \mu m g \qF(\qe), \label{eq:motion_vel}\\
\frac{1}{2} m R^2 \frac{d \qo}{d t} & = & - \mu m g R \qT(\qe).
\label{eq:motion_omega}
\end{eqnarray}
By introducing dimensionless velocities and time as
$\qvv = \qv / \sqrt{R g} \mu$,
$\qoo = \qo \sqrt{R / g} / \mu$, and 
$t^* = t \sqrt{g / R}$,
Eqs.\ (\ref{eq:motion_vel}) and (\ref{eq:motion_omega}) reduce to
\begin{eqnarray}
\frac{d \qvv}{d t^*} & = & - \qF(\qe), \label{eq:motion_velb}\\
\frac{d \qoo}{d t^*} & = & - 2 \qT(\qe) \label{eq:motion_omegab}
\end{eqnarray}
with $\qe = \qvv / \qoo$.  As $\qF(\qe)$ and $\qT(\qe)$ are positive
for $\qe > 0$, the translational and angular velocities are strictly
monotonically decreasing in time, as expected.  Now the question
arises: Is it possible that any of them reaches zero before the other,
i.e., may it happen that an initially sliding and spinning disk after
some time is only sliding or spinning?  We solved Eqs.\
(\ref{eq:motion_velb}) and (\ref{eq:motion_omegab}) numerically with
many different initial conditions. The results indicate, as can be
seen in Fig.~\ref{fig:trajectories}(a), that $\qvv$ and $\qoo$ always
reach zero together, meaning that \textsl{the disk always stops its
sliding and spinning motion at the same moment}.
Before we show its proof, let us try to explain qualitatively what is
happening. If the velocity is much higher than the angular velocity
($\qv \gg R \qo$, i.e., $\qe \gg 1$), then the friction torque is
negligible compared to the force, see
Fig.~\ref{fig:F_T_e}(a). Therefore, the velocity decreases with a higher
rate than the angular velocity, and $\qe$ decreases.  On the other
hand, if the angular velocity is much higher than the velocity ($\qe
\ll 1$), then the friction torque is higher than the force, and $\qe$
increases. Thus a negative feedback effectively equilibrates the
sliding and spinning motion.  Indeed, the numerical results show this
behavior, as $\qe$ always tends to the same value,
$\qe_{0}\approx 0.653$,
when the motion stops [Fig.~\ref{fig:trajectories}(b)]. This means 
that $\qoo$ and $\qvv$ not only reach zero simultaneously, but also that their
ratio approaches a universal value, irrespective of the initial conditions. 

\begin{figure}
\centerline{\epsfig{figure=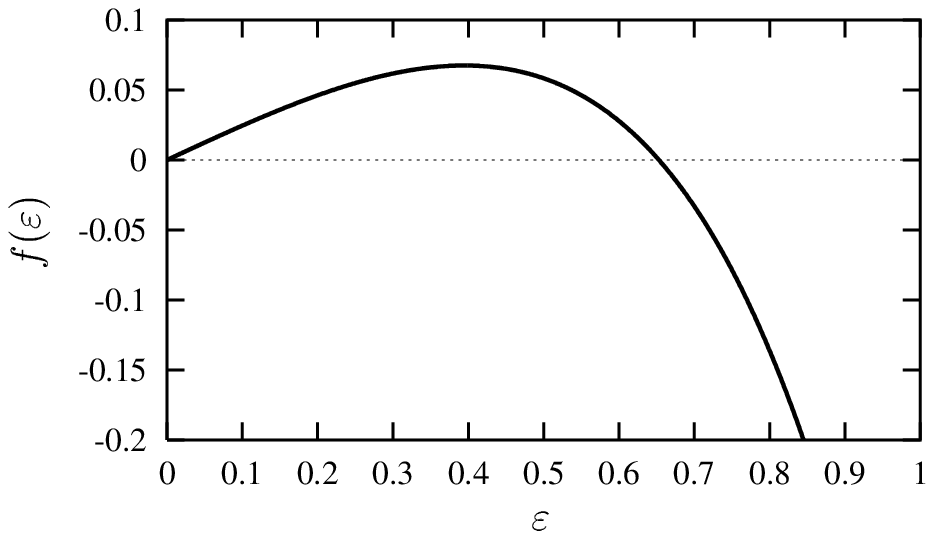,width=7.7cm}}
\caption{Function $f(\qe) = \qe - \qF(\qe)/2\qT(\qe)$,
the r.h.s.\ of differential equation (\ref{eq:dedx}).
It has zero value at $\qe = 0$ and $\qe_{0} \approx 0.653$,
and is positive for $0 < \qe < \qe_0$, negative for $\qe_0 < \qe$.}
\label{fig:rhs}
\end{figure}

In order to prove that $\qe$ always has this  
value at the end of the motion, we derive an autonomous differential
equation for $\qe$ from equations (\ref{eq:motion_velb}) and
(\ref{eq:motion_omegab}) using the variable transformation $x = - \ln
\qoo$:
\begin{equation}
\frac{d\qe}{dx} = \qe - \frac{\qF(\qe)}{2\qT(\qe)} \equiv f(\qe).
\label{eq:dedx}
\end{equation}
Note that $\qoo \ra 0$, the condition of stopping, now corresponds to
$x \ra \infty$ (with the exception of pure sliding motion). For small
$\qe$ the right hand side of Eq.\ (\ref{eq:dedx}) vanishes like
$f(\qe) \approx \qe/4$, while it behaves asymptotically for $\qe \ra
\infty$ like $f(\qe) \approx - \qe$.  In between, at $\qe_{0}$, it
changes sign (Fig.~\ref{fig:rhs}).
Therefore Eq.\ (\ref{eq:dedx}) has three fixed points: Two of them,
$\qe = 0$ and $\qe = \infty$, are trivial and correspond to pure
spinning or pure sliding motion, respectively.  For all other initial
conditions ($0 < \qe < \infty$), corresponding to initial sliding
\textsl{and} spinning, $\qe_{0}$ is the attractive fixed point,
meaning that $\qe$ has this value just before the disk stops its
motion, which is what we wanted to prove.

\begin{figure}
\centerline{\epsfig{figure=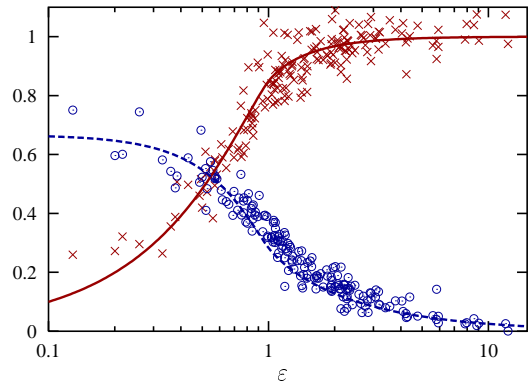,width=7.7cm}}
\caption{Experimental data: the instantaneous dimensionless friction
force (crosses) and torque (circles), acting on a sliding and spinning
CD disk, as functions of $\qe$. One pair of data points is presented
for each throw, taken shortly after setting the disk into motion.  The
corresponding theoretical curves are also displayed: $\qF(\qe)$ (solid
line) and $\qT(\qe)$ (dashed line).}
\label{fig:experiment}
\end{figure}

We also performed a simple experiment to measure the friction force
and torque acting on a sliding and spinning disk. We set a standard
writable CD disk ($R = 6\,\mbox{cm}$), with its data carrying side
down, into motion manually on a horizontal polyamid fabric surface
several times, and recorded its motion with a Sony DCR-VX2000E PAL
digital video camera (25 images/second).  Then we processed the images
to obtain the position and the orientation of the disk as a function
of time. We fitted the position and angular data with second degree
polynomials to get the instantaneous translational and angular velocity and
acceleration. Then, assuming that only gravitational and frictional
forces were acting on the disk, using Eqs.\ (\ref{eq:motion_vel})
and (\ref{eq:motion_omega}), and having only the friction coefficient
as a fit parameter, we were able to plot functions $\qF(\qe)$ and
$\qT(\qe)$, see Fig.~\ref{fig:experiment}. We had, however, a minor
complication: We observed that the friction coefficient slightly
increased linearly with the number of throws, $n$. We found that $\mu
= 0.202 + 0.00053 n$ was a good fit, and the data presented in
Fig.~\ref{fig:experiment} was obtained using this ``time-dependent''
friction coefficient in the data processing. Investigating the disk we
concluded that the reason for the increase of $\mu$ was probably that the film
on the CD was gradually removed.  The disk also had a $1.5\,\mbox{cm}$
diameter hole at its center, but a numerical calculation of $\qF(\qe)$
and $\qT(\qe)$ for this geometry showed that deviation from the full
disk case is not significant, it is much smaller than the accuracy of
the measured data. Our experimental data also showed that the disk
always stopped its sliding and spinning motion at the same moment
(within error).

We used the sliding and spinning disk as a simple, illustrative
example to show how friction force and torque are coupled. In this
case we were able to derive all results analytically, because the
local pressure is everywhere the same in the contact area. However, in
general the pressure distribution over the contact area will be
non-uniform. As an example, if we replace the flat disk by a cylinder
standing on one of its flat faces, then the friction force leads to a
torque with respect to the center of mass. Provided that the cylinder
does not topple, this torque must be compensated by a pressure
increase at the front and a pressure decrease at the rear part of the
contact area. Therefore the spinning motion induces a friction
component perpendicular to the translational motion, in the direction
of $\qvel \times \qomega$. Hence, in contrast to the straight sliding of a
flat spinning disk, the path of the cylinder will be curved in this
direction. This resembles the Magnus effect, although the physical
origin is completely different.

The pressure distribution can also depend on the shape and elastic
properties of the sliding body. For instance, if it is a sphere,
linear elasticity theory predicts a $\sqrt{1 - r^2/R^2}$ shaped radial
pressure function \cite{Landau}.  We calculated the $\qF(\qe)$ and
$\qT(\qe)$ curves numerically for this case and found that their
qualitative behavior remains the same \cite{footnote:rolling}.
Therefore coupling between the friction force and torque is still
present: For large $\qe$ torsion friction is suppressed by sliding,
for small $\qe$ sliding friction gets reduced by spinning. This may
explain, why the translational motion of a fast spinning top is hardly
decelerated.

Now let us consider what impact the coupling between sliding and
spinning motion may have on the physics of granular media. In one
limiting case, when the particles are spherical and very hard, torsion
friction typically can be neglected, as the contact area is very
small, hence $\qe \gg 1$. However, on the one hand, real particles are
usually non-spherical, or they may be very soft, so that the size of
the contact area can be comparable to that of the particle. On the
other hand, even in the case of hard spherical particles with contact
radius $r$ much smaller than the particle radius $R$, the coupling
between sliding and spinning motion can have subtle consequences. As
the torsion friction is very small in this case, one can expect that
the spinning degree of freedom can be easily excited. Typical sliding
velocities will be comparable to $\omega_{\rm n} R$, so that $\qe
\approx R/r \gg 1$. In this case, the sliding friction is basically
$\mu \qFn$. However, when a granular packing relaxes into a static
configuration, the coupling between sliding and spinning becomes
important. For example, if the initial sliding velocity is zero, but
the spinning degree of freedom is excited (i.e., $\qe=0$), \textsl{an
arbitrarily small force can induce sliding}.

An important extension of this work will be to investigate 
the coupling between the {\em static} friction force and
torque. As static friction is different from sliding friction, and its
theory is somewhat more difficult, we cannot expect that the maximum
static friction forces and torques will lie on the friction-torque
curve in Fig.~\ref{fig:F_T_e}(b).  However, we have preliminary
evidence that such coupling is present also in the static case.  We
expect that the threshold torques and forces needed to turn a sticking
contact into a sliding and/or spinning one form a curve which lies
above the one in Fig.~\ref{fig:F_T_e}(b). In particular, this implies
that the application of a torque at a sticking contact makes it easier
to excite the sliding degree of freedom. For hard spherical spheres a
very small torque will already have this effect.

Finally, the fact that $\qF(\qe)$ and $\qT(\qe)$ depend on the
pressure distribution in the contact area raises the question, whether
the ``inverse problem'' has a unique solution, i.e., are the
experimentally accessible functions $\qF(\qe)$ and $\qT(\qe)$ a
fingerprint of the pressure distribution in the contact area? This
would also be an interesting question to study in the future.

We would like to thank J\'anos Kert\'esz, P\'eter Gn\"adig, Dirk
Kadau, and Lothar Brendel for useful discussions, and Detlef
Wildenberg (Audio-Visual Media Center, Gerhard-Mercator University)
for his help to video record the experiment.  We acknowledge funding
from DFG Graduate College 277 (G.B.), DFG Grant Wo577/3-1 (Z.F.), and
DAAD (T.U.).


\begin{thebibliography}{99}
\bibitem{footnote:rolling} In the case of a sphere, also rolling
  should be taken into account, leading to further complications.
  However, for the sake of simplicity, here we neglected rolling.
\end{thebibliography}
\end{document}